\documentclass[12pt]{article}

\def\BEq{\begin{equation}}
\def\EEq{\end{equation}}
\def\BEqA{\begin{eqnarray}}
\def\EEqA{\end{eqnarray}}
\def\BEn{\begin{enumerate}}
\def\EEn{\end{enumerate}}

\def\a{\alpha}

\def\b{\beta}

\def\d{\delta}

\def\s{\sigma}

\def\dag{\dagger}

\def\adj{{^{\dag}}}

\usepackage{makeidx,graphics,latexsym}

\begin{document}

\title{{\bf Generation of high-fidelity controlled-not logic
gates by coupled superconducting qubits}}

\author{
Andrei
Galiautdinov
\\
\normalsize{\it Department of Physics and Astronomy,
University of Georgia, Athens, Georgia
30602, U.S.A.}\\
\small{ag@physast.uga.edu}
}

\date{\today}

\maketitle

\abstract{Building on the previous results of the Weyl chamber
steering method, we demonstrate how to generate high-fidelity
controlled-not by direct application of certain, physically relevant
Hamiltonians with {\it fixed coupling constants} containing
Rabi terms. Such Hamiltonians are often used to describe two
superconducting qubits driven by local {\it rf}-pulses.

It is found that in order to achieve 100\% fidelity in a system
with {\it capacitive} coupling of strength $g$ one Rabi term suffices.
We give the exact values of the physical parameters needed to implement
such CNOT. The gate time and all possible Rabi frequencies
are found to be $t=\pi/(2g)$ and
$\Omega_1/g = \sqrt{64n^2-1},\; n =1,2,3,\dots$.
Generation of a perfect CNOT in a system with {\it inductive}
coupling, characterized by additional constant $k$, requires the
presence of both Rabi terms. The gate time is again $t = \pi/(2g)$,
but now there is an infinite number of solutions, each of which is
valid in a certain range of $k$ and is characterized by a pair of positive
integers $(n,m)$,
\[
\frac{\Omega_{1,2}}{g} =
\sqrt{ 16n^2-\left(\frac{k-1}{2}\right)^2}\pm 
\sqrt{ 16m^2-\left(\frac{k+1}{2}\right)^2}.
\]
We distinguish two cases, depending on the sign of the coupling constant:
(1) the {\it antiferromagnetic} case ($k\geq 0$) with $n\geq m = 0, 1,2,\dots$;
(2) the {\it ferromagnetic} case ($k\leq 0$) with $n> m = 0, 1, 2, \dots$.

The paper concludes with consideration of fidelity degradation by switching
to resonance. Simulation of time-evolution based on the 4th order Magnus
expansion reveals characteristics of the gate similar to those found in
the exact case, with slightly shorter gate time and shifted values of
Rabi frequencies. 
}

PACS number(s): 03.67.Lx, 03.65.Fd, 85.25.-j

\newpage


\section{Introduction}

The goal of this paper is to show how to successfully generate, with
a minimal amount of effort, a high-fidelity CNOT in systems consisting
of coupled superconducting qubits described by Hamiltonians with
{\it coupling constants fixed by architecture}. We want to find out if
making a CNOT in such systems can, at least theoretically, be reduced
to a simple push of a button. Once initially set, the system itself
would then take care of the needed gate by freely evolving from the
identity of the unitary group to the target along the corresponding geodesic.

We thus approach the subject of time-control from the ``opposite'',
non-optimal end. The plug-and-play implementations considered in this paper
may be useful for simple gate designs which do not rely on sophisticated
time-control of their physical parameters. The question of simplicity will
also become important once the best qubit is identified, and put together
with thousands of others to form an integrated circuit. Optimally
controlling \cite{KHANEJAPRA63} qubits in such a circuit would become
a difficult task, potentially leading to a compromise between the optimal
and the simple. Thus, in this paper we limit our attention mostly to
time-independent Hamiltonians. The only place where time dependence
necessarily arises is in the analysis of fidelity degradation due to
switching to resonance, considered in Section \ref{sec:FASTSWITCHING}.

Two important developments motivated our work. The first was a series
of experiments  \cite{CLARKEETAL88, JOHNSONETAL2003, DEVORET04} which
showed that the macroscopic quantum states \cite{LEGGETT80, CALDEIRALEGGETT83}
of Josephson junctions have coherence times long enough for such states to
be used as qubits. That led to our choice of the physical Hamiltonians. 
The other development was the work on the Weyl chamber steering method
\cite{ZHANG, ZHANG1} in which several examples of direct generation of
various important gates, including CNOT, was considered from a purely
geometrical standpoint. 

In what follows we will use the square brackets [$\cdots$] to denote
the gate --- a representative of a given local class --- located in
a Weyl chamber.

\section{The concept of local equivalence}
\label{sec:LOCALEQUIVALENCE}

We begin by recalling the notion of local equivalence within the special
unitary group $SU(4)$. Two gates $U_1, V_1\in SU(4)$ are {\it locally
equivalent} if they differ only by local transformations:
\BEq
U_1\sim V_1 \Longleftrightarrow U_1 = k_1\, V_1\, k_2,
\quad k_1, k_2 \in SU(2)\otimes SU(2).
\EEq

It is obvious that local equivalence is an example of equivalence
relation: it is reflexive ($U\sim U$), symmetric
($U\sim V \Longleftrightarrow V\sim U$), and transitive
($U\sim V, \; V \sim W \Longrightarrow U\sim W$). When an equivalence
relation is given on a set, it partitions the set into disjoint classes.
Thus the entire group $SU(4)$ is partitioned into classes of locally
equivalent elements; any such element (a two qubit gate) belongs to
one and only one local equivalence class.

To generalize the concept of local equivalence to the full $U(4)$ we
use the fact that any $U\in U(4)$ can be written in the form
$U=e^{i\, \a}\,U_1$, with $U_1\in SU(4)$ (even though, as groups,
$U(4)\neq U(1)\otimes SU(4)$). Then two gates
$U = e^{i\,\a}\,U_1, V=e^{i\,\b}\,V_1 \in U(4)$ are locally equivalent
if their representatives in $SU(4)$ are equivalent:
\BEq
U \sim V \Longleftrightarrow U_1 \sim V_1.
\EEq

One way to represent local classes in $SU(4)$ is with the help of a
Cartan decomposition of the underlying Lie algebra $L = su(4)$. This
can be seen as follows:

For any {\it semi-simple} Lie algebra $L$, its Cartan decomposition
is a direct sum of two {\it subspaces} $L = K\oplus P$, one of which,
$K$, is itself a Lie algebra, $[K,\, K] \subset K$, and the other, $P$,
satisfies $[P,\, P]\subset K$, $[P, \, K] \subset P$. Because
$K\cap P = 0$, any subalgebra $A\subset  P$ must necessarily be abelian.
Such {\it maximal} abelian subalgebra $A_C \subset P$ is called the Cartan
subalgebra of $L$ relative to the given Cartan decomposition. An important
Cartan decomposition theorem (for {\it groups}) states that if a semisimple
algebra $L$ has decomposition $L = K\oplus P$, then any element
$U_1 \in \exp L$ can always be written in the form 
\BEq
U_1 = k_1 \, V_1 \, k_2,
\EEq
where $V_1 \in \exp A_C$, and $k_1, k_2 \in \exp K $.

An example of a Cartan decomposition is provided by $su(2)$,
for which we can take $K = {\rm span}\{\frac{i}{2} \s_z\}$,  
$P = {\rm span}\{\frac{i}{2} \s_x, \; \frac{i}{2} \s_y\}$, and
$A_C = {\rm span}\{\frac{i}{2} \s_x\}$.

Cartan decompositions of $su(4)$ are more complicated, but a special
choice \cite{KHANEJAPRA63} of $K$ spanned by the local operators
\BEq
X_1=\frac{i}{2} \s_x^1, \quad Y_1=\frac{i}{2} \s_y^1,\quad 
Z_1=\frac{i}{2} \s_z^1,\quad X_2=\frac{i}{2}\s_x^2, \quad 
Y_2=\frac{i}{2}\s_y^2, \quad Z_2=\frac{i}{2}\s_z^2,
\EEq
and $P$ spanned by all nonlocal operators
$\frac{i}{2}\,\s_\a^1\s_\b^2$ ($\a,\b = x, y, z$) is particularly 
useful. If we now take the span of mutually commuting operators
\BEq
XX=\frac{i}{2}\,\s_x^1\s_x^2, \quad 
YY=\frac{i}{2}\,\s_y^1\s_y^2,\quad  ZZ=\frac{i}{2}\,\s_z^1\s_z^2
\EEq
to serve as $A_C$, the Cartan decomposition theorem applied 
to $SU(4) = \exp [su(4)]$ gives
\BEq
\label{eq:CARTADECOMPOSITIONsu4}
U_1 = k_1 \, \exp\left(c_1 XX + c_2 YY +c_3 ZZ \right) k_2.
\EEq
This results in a convenient representation of $SU(4)$ local 
equivalence classes by the {\it class vectors} $[c_1, c_2, c_3]\in A_C$.
Representation of the $U(4)$ classes is up-to-phase the same,
\BEq
U = e^{i\,\a}\, k_1 \, \exp\left(c_1 XX + c_2 YY +c_3 ZZ \right) \, k_2.
\EEq

The correspondence between class vectors and equivalence classes is 
{\it not} one-to-one: there are certain symmetries which map class 
vectors to other class vectors of the same equivalence class. However it was
shown in \cite{ZHANG} that the correspondence can be made unique by 
restricting class vectors to a tetrahedral region in $A_C$, called a 
Weyl chamber. One such chamber, denoted by $a^+$, is chosen to be 
canonical --- it is described by the following three conditions 
\cite{ZHANG, TUCCI}:
\begin{itemize}
\item $\pi > c_1 \geq c_2 \geq c_3 \geq 0$,
\item $c_1+c_2 \leq \pi$,
\item if $c_3 =0$, then $c_1\leq \pi/2$.
\end{itemize}
Even though the choice of the Weyl chamber is not unique, when fixed, 
it becomes a powerful tool for analyzing time evolution on the full $U(4)$.

In the original papers on the subject \cite{ZHANG, ZHANG1}, several 
implementations of various useful gates by steering on $a^+$ have been 
demonstrated. Of particular interest to us is the way [CNOT] = 
$[\pi/2, 0, 0]\in a^+$ was generated by a {\it single} application 
of Hamiltonians with {\it controllable} Ising type interaction, such as
\BEq
H_{yy}=\Omega_{1x}\s_x^1 + \Omega_{2x}\s_x^2 + \Omega_{1z}\s_z^1 + 
\Omega_{2z}\s_z^2 + g \s_y^1\s_y^2, 
\EEq
and similarly for $H_{xx}, H_{zz}$. 
In \cite{ZHANG, ZHANG1}, the local invariants of the corresponding 
unitary evolution $U(t)=\exp(iH_{yy}t)$ were first found in closed 
analytic form. Numerical analysis then produced the values for the 
time and the coupling constants needed to achieve a perfect [CNOT] 
in just one application. The $yy$ and $zz$ cases are particularly 
interesting because their interaction parts (when viewed as elements 
of $A_C$) point in the direction ``perpendicular'' to [CNOT].

Let us now turn to Hamiltonians used to describe coupled superconducting 
qubits. Such Hamiltonians may contain several Ising terms, which we 
choose to be fixed by architecture ({\it cf.} \cite{PLOURDE, KOFMAN}). 
Our goal is to find out if they are also capable of generating [CNOT] 
in a single application, and if so, how many Rabi terms are required 
in order for such generation to be successful.

\section{Hamiltonians for coupled superconducting qubits}
\label{sec:HAMILTONIAN}

When restricted to the two-qubit subspace, the Hamiltonian for 
two capacitively coupled phase qubits driven by {\it rf}-pulses 
is given by
\BEq
H_{\rm fast} = \underbrace{-\sum_{i=1}^2\frac{\omega_i}{2}\s_z^i}_{H_0} 
+ \underbrace{\sum_{i=1}^2\Omega_i\cos(\omega_{\rm rf}^i t
+\d_i)\s_x^i
+ g \s_y^1\s_y^2}_{V}, \quad g>0.
\EEq
The level spacings $\omega_i$ are tunable between about 7 to 10 GHz. 
The Rabi frequencies $\Omega_i$ and the phases $\d_i$ are fully controllable.

As it stands, this Hamiltonian is not suitable for actual gate 
implementations because of the very small time scale ($t_{\rm sc}\sim 0.1$ ns) 
associated with $\omega_i$. The scale roughly corresponds to the time it 
takes to make a  ``round trip'' on $SU(4)$ when going in the direction 
of $-iH\in su(4)$. To attain high fidelity of the gate we would have to 
control the pulses with accuracy which is experimentally unattainable.

To avoid this problem, the rotating wave approximation (RWA) is used. 
We set the frequency of the external pulse to be equal to the transition 
frequency of the system (resonance), 
$\omega_{\rm rf}^i = \omega_i \equiv \omega$, bring the phases to zero, 
and work in the interaction picture, in which the time evolution is governed by
\BEqA
H &=& e^{iH_0t}Ve^{-iH_0t} \nonumber \\
&=& \sum_{i=1}^2\Omega_i\left[ \cos^2(\omega t) \s_x^i 
+  \frac{\sin(2\omega t)}{2}\s_y^i \right]
+g\left[\cos(\omega t)\s_y^1-\sin(\omega t)\s_x^1\right] 
\left[\cos(\omega t)\s_y^2-\sin(\omega t)\s_x^2\right]\;.
\nonumber \\
\EEqA
After averaging over fast oscillations the Hamiltonian becomes
\BEq
H = \frac{\Omega_1}{2}\s_x^1 + \frac{\Omega_2}{2}\s_x^2
	+ \frac{g}{2}\left(\s_x^1\s_x^2+\s_y^1\s_y^2\right),
\EEq
or in the language of Lie algebra $su(4)$,
\BEq
\label{eq:H}
-iHt = -t\, \left[\Omega_1\,X_1 + \Omega_2\,X_2
	+g\left(XX+YY\right)\right].
\EEq

There is an interesting modification of this qubit architecture based 
on inductive coupling between the qubits (see Appendix). In that case an additional 
coupling constant $k$ is introduced resulting in the full Hamiltonian
\BEq
\label{eq:tildeH}
-i Ht = -t\, \left[\Omega_1\,X_1 + \Omega_2\,X_2
	+g\left(XX+YY+kZZ\right)\right].
\EEq

\section{Implementing [CNOT] by time-independent Hamiltonians}
\label{sec:EXISTENCE}

\subsection{The $XX+YY$ case}
\label{sec:EXISTENCEXXYY}

Here we show how to implement the perfect [CNOT] by a single application 
of the experimentally available Hamiltonian (\ref{eq:H}) with one Rabi term
\BEq
\label{eq:H1Rabi}
-iHt = -t\, \left[\Omega_1\,X_1 
	+g\left(XX+YY\right)\right]\;.
\EEq

We first establish a useful group-theoretical result about the exponentials 
of (\ref{eq:H1Rabi}). We consider a subalgebra of $su(4)$, 
$A_0 = {\rm span}\{X_1, XX, YY, ZY\}$, whose generators obey the 
commutation relations summarized in the following table:
\begin{equation}
\label{eq:COMMTABLE}
A_0: \quad
 \begin{array}{|c|cccc|}\hline
    & X_1 & XX & YY  & ZY  \\ \hline
X_1 &  0  &  0 & -ZY & YY  \\ 
XX  &  0  &  0 &   0 &   0 \\ 
YY  &  ZY &  0 &   0 &   -X_1\\
ZY  & -YY &  0 &  X_1&   0\\ \hline
 \end{array}\quad .
\end{equation}
Notice that $ZY$ generator had to be added to the set of generators 
in order to ensure closure under commutation.

The generator $XX$ is central in $A_0$. Its span is a one-dimensional 
abelian ideal $I\subset A_0$, which makes $A_0$ into a 
{\it non}-semisimple Lie algebra. This algebra is a direct sum of two subspaces,
$A_0 = I \oplus A_1$, with $A_1 = {\rm span}\{X_1, YY, ZY\}$ 
{\it isomorphic} to $su(2)$. 

 Exponentiation of $A_0$ results in a {\it four}-dimensional 
 subgroup $G_0=\exp A_0\subset SU(4)$ whose every element $W$ 
 can be written as a product
 \BEq
 \label{eq:W}
  W = e^{ - c_1\, XX } 
  \underbrace{e^{ -a\, X_1 -c_2\, YY -b\, ZY}}_{{\rm an\; element\; of\;} \exp A_1}. 
 \EEq
 Because $A_1$ has its own Cartan decomposition ({\it e.g.},
 $A_1 = {\rm span}\{X_1\} \oplus  {\rm span} \{YY, ZY\}$ with Cartan subalgebra
 $A_C = {\rm span}\{YY\}$), application of the Cartan decomposition theorem 
 to the group  $G_1=\exp A_1$ gives  
 \BEq
  W = e^{-c_1\, XX}\,e^{ -\a\, X_1}\,e^{ -c_2\, YY}\, e^{ -\b\, X_1}. 
 \EEq

Now, because $e^{-iHt}=e^{-t[\Omega_1 X_1 + g(XX+YY)]}$ is itself an element 
of $G_0=\exp A_0$, it can also be written as such product
\BEq
\label{eq:CARTANDECOMPRESTRICTEDXXYY}
e^{-t[\Omega_1 X_1 + g(XX+YY)]}= 
e^{ -\a \, X_1}\, e^{- c_1\, XX-c_2 \, YY}\, e^{-\b \, X_1}\,,
\EEq
where the commutativity of $X_1$ and $XX$ has been used to rearrange the 
factors on the right.

Thus given any gate $(t, \Omega_1, g)$ generated by $H$ in (\ref{eq:H1Rabi}), 
with coupling $g$ fixed by architecture, there always exists a set of four 
numbers $(\a, \, c_1, \, c_2,\, \b)$ which represent that gate exactly. 
This is essentially the same good old Cartan decomposition theorem 
(\ref{eq:CARTADECOMPOSITIONsu4}) for $SU(4)$, restricted to a physically 
relevant subgroup $G_0$ associated with experimentally available Hamiltonian 
(\ref{eq:H1Rabi}).

By fixing $\Omega_1$ and evolving the system freely under the action of 
its Hamiltonian, we generate a flow along a straight line in $su(4)$ which 
has its image in both, the Weyl chamber and the group $SU(4)$.

The converse is unfortunately not necessarily true: for a given gate 
$(\a, \, c_1, \, c_2,\, \b)$ it is not always possible to find $(t, \Omega_1)$ 
generating that gate. This is because of noncommutativity between $X_1$ and $YY$.

Even though there is no guarantee that $H$ would be able to reach every 
$(c_1,\,c_2, 0)$ equivalence class in a single application, it is 
interesting to find out what actually {\it can} be reached. In particular, 
we want to know if [CNOT] is reachable.

It is obvious that if $\Omega_1 =0$, then any local class with $c_1=c_2$ 
and  $c_3=0$ can be directly reached. What if $\Omega_1 \neq 0$?

We will be interested in purely geodesic solutions to 
(\ref{eq:CARTANDECOMPRESTRICTEDXXYY}). Thus, setting $\a = \b = 0$ and 
directly exponentiating in 
(\ref{eq:CARTANDECOMPRESTRICTEDXXYY}) we get the matrix equation
\BEq
\left[
\begin{array}{cccc}
\cos b_1 - \cos a & 0 & i cd \sin a & i(\sin b_1 - d \sin a) 	\\
0 & \cos b_2-\cos a &i(\sin b_2 + d\sin a) & icd \sin a 			\\
icd\sin a & i(\sin b_2 + d \sin a) & \cos b_2 - \cos a & 0			\\
i(\sin b_1 - d \sin a) & icd\sin a & 0 & \cos b_1 - \cos a			\\
\end{array}
\right]=
0,
\EEq
where
\BEq
a=\frac{tg}{2}\, \sqrt{1+\left(\frac{\Omega_1}{g}\right)^2}, \quad
b_{1,2} = \frac{1}{2}( tg - c_1 \pm c_2), \quad
c=\frac{\Omega_1}{g},\quad
d=\frac{1}{\sqrt{1+\left(\frac{\Omega_1}{g}\right)^2}}.
\EEq
If $\Omega_1 \neq 0$, then $\sin a = 0$, leading to $\cos a = \pm 1$ 
and $\cos b_{1} = \cos b_2 = \pm 1$, $\sin b_{1,2} =0$.
This is only possible if $b_1 = b_2 + 2\pi m$, or
\BEq
c_2 = 2\pi m, \quad m = 0, 1, 2, \dots .
\EEq
The minimal time solution is therefore
\BEqA
\label{eq:XXYYSOLUTION}
t_{\rm min} &=& \frac{c_1}{g}, \nonumber \\
\frac{\Omega_1}{g}&=& \sqrt{\left( \frac{4\pi n}{c_1}\right)^2-1}, 
\quad n = 1,2,3,\dots .
\EEqA
Thus, by following $H$ along its geodesic for a time $t_{\rm min}$ we 
can (exactly) reach any 
equivalence class located on the $XX$ axis of the Weyl chamber. It 
is clear that the coupling constant $g$ plays the role of the scaling 
factor between the coordinate $c_1$ of the point on this axis and the 
actual, {\it physical} time required to reach that point.

We want to emphasize that the corresponding trajectory in the Weyl 
chamber is {\it not} a straight line from $O$ to [CNOT], even though 
is is a straight line in the full Lie algebra $su(4)$. This is because 
with $\Omega_{1}$ now being fixed by (\ref{eq:XXYYSOLUTION}), it is not 
true that $e^{-t[\Omega_1 X_1 + g(XX+YY)]} \sim e^{- c_1(t)\, XX}$ is 
satisfied for all intermediate times $0< t < t_{\rm min}$. In general 
for such times, $c_2(t)\neq 0$ in (\ref{eq:CARTANDECOMPRESTRICTEDXXYY}).

One important local class belonging to the $XX$ axis is [CNOT], which 
can be made by exponentiating the Hamiltonian $-iH_{\rm [CNOT]}t=-tXX$ 
with $t=\pi/2$. The corresponding matrix in the computational basis is
\BEqA
\label{eq:CNOTMATRIX}
{\rm [CNOT]} &=& \exp\left( -\frac{\pi}{2}\, XX \right) = \frac{1}{\sqrt{2}}
\left[
\begin{array}{cccc}
1&0&0&-i\\
0&1&-i&0\\
0&-i&1&0\\
-i&0&0&1\\
\end{array}
\right]\in SU(4).
\EEqA
The Makhlin invariants \cite{MAKHLIN} of this [CNOT] are $G_1=0$ 
and $G_2=1$, as required, so this is a perfectly acceptable [CNOT] 
out of which the canonical CNOT with matrix
\BEqA
\label{eq:trueCNOTMATRIX}
{\rm CNOT} &=& 
\left[
\begin{array}{cccc}
1&0&0&0\\
0&1&0&0\\
0&0&0&1\\
0&0&1&0\\
\end{array}
\right]\in U(4)
\EEqA
can be produced by local manipulation of individual qubits. We will 
give some examples of the needed for this local {\it rf}-pulses in 
Section \ref{sec:rfpulses}.

Thus, setting $c_1 = \pi/2$ in (\ref{eq:XXYYSOLUTION}), we get
\BEqA
t_{\rm min} &=& \pi/(2g), \nonumber \\
\Omega_1/g&=& \sqrt{64n^2-1}, \quad n = 1,2,3,\dots \;.
\EEqA
The simplest exact [CNOT] is then 
\BEq
 {\rm [CNOT]} =  e^{-\frac{\pi}{2g}\,[\sqrt{63}g X_1 + g(XX+YY)]}\;.
\EEq

For completeness we mention another [CNOT] solution to 
(\ref{eq:CARTANDECOMPRESTRICTEDXXYY}),
\BEq
\label{eq:UPTOPHASESOLUTION}
\Omega_1/g= \sqrt{16n^2-1}, \quad n = 1,2,3,\dots \;,
\EEq
which is {\it up-to-phase} geodesic,
\BEq
 {\rm [CNOT]} =  e^{2\pi \, X_1} e^{-\frac{\pi}{2g}\,
 [\Omega_1 X_1 + g(XX+YY)]}\;.
\EEq

\subsection{The $XX+YY+kZZ$ case}

We are now going to show that a single application of the 
Hamiltonian given in (\ref{eq:tildeH}) generates the exact 
[CNOT], provided both Rabi terms are kept.

Derivation parallels the previous case.
We concentrate on subalgebra $A_0\subset su(4)$:
\begin{equation}
\label{eq:COMMTABLEk}
A_0: \quad
 \begin{array}{|c|ccccccc|}\hline
    & X_1 & X_2 & XX& YY & ZZ &  YZ		&		ZY \\ \hline
X_1 &  0  &  0  & 0 & -ZY& YZ &  -ZZ	&		YY  \\
X_2 &  0  &  0  & 0 & -YZ& ZY & 	YY	&	-ZZ\\ 
XX  &  0  &  0  & 0 &  0 & 0  &    0	&	0\\ 
YY &    ZY&  YZ & 0 &  0 & 0  &		-X_2&-X_1\\ 
ZZ &  -YZ  & -ZY& 0 &  0 & 0  & 	X_1	&X_2\\ 
YZ &  ZZ   & -YY& 0 & X_1&-X_1&  		0	&	0 \\ 
ZY & -YY   &  ZZ& 0 & X_1&-X_2&  		0	&	0 \\ \hline
 \end{array}\quad .
\end{equation}

The generator $XX$ is again central in $A_0$, spanning a 
one-dimensional abelian ideal $I\subset A_0$. This makes 
$A_0$ into a non-semisimple Lie algebra which is a direct 
sum of two subspaces,
$A_0 = I \oplus A_1$, where $A_1 = {\rm span}\{X_1, X_2, YY, ZZ, YZ, ZY\}$.

 Exponentiation of $A_0$ results in a {\it seven}-dimensional 
 subgroup $G_0=\exp A_0\subset SU(4)$ whose every element $W$ 
 can be written as a product
 \BEq
 \label{eq:Wk}
  W = e^{ - c_1\, XX } 
  \underbrace{e^{ -a\, X_1 - b\, X_2 - c_2\,YY -c_3\,ZZ
  - d\,YZ - f \, ZY}}_{{\rm an\; element\; of\;} \exp A_1}. 
 \EEq
 Because $A_1$ has its own Cartan decomposition ({\it e.g.},
 $A_1 = {\rm span}\{X_1, X_2\} \,\oplus \, {\rm span} \{YY, ZZ, YZ, ZY\}$ 
 with a Cartan subalgebra
 $A_C = {\rm span}\{YY, ZZ\}$), application of the Cartan decomposition 
 theorem to the group  $G_1=\exp A_1$ gives  
 \BEq
  W = e^{-c_1\, XX}\,e^{ -\a\, X_1 -\b\,X_2}\,e^{ -c_2\, YY -c_3 \,ZZ}\, 
  e^{ -\eta \, X_1 - \xi \, X_2}. 
 \EEq

Since $e^{-i Ht}=e^{-t[\Omega_1 X_1 +\Omega_2 X_2 - g(XX+YY+kZZ)]}$ is 
itself an element of $G_0=\exp A_0$, it can also be written as such product
\BEq
\label{eq:kZZtimeevolution}
e^{-t[\Omega_1 X_1 + \Omega_2 X_2 + g(XX+YY+kZZ)]}=e^{ -\a\, X_1 -\b\,X_2}\,
e^{ -c_1\, XX -c_2\, YY -c_3 \,ZZ}\, e^{ -\eta \, X_1 - \xi \, X_2}.
\EEq

Thus, given any gate $(t, \Omega_1, \Omega_2, g, k)$ generated by $ H$ 
in (\ref{eq:tildeH}), with couplings $g$ and $k$ fixed by architecture, 
there always exists a set of seven numbers 
$(\a, \, \b, \,  c_1, \, c_2,\,c_3,\, \eta, \, \xi)$ which represent 
that gate exactly. This is again the Cartan decomposition theorem for 
$SU(4)$, this time restricted to a physically relevant subgroup $G_0$ 
associated with experimentally available Hamiltonian (\ref{eq:tildeH}).

With $\a=\b = \eta =\xi =0$, equation (\ref{eq:kZZtimeevolution}) has 
an infinite number of minimal time solutions labelled by pairs of 
integers $(n,m)$ and valid in certain
intervals of $k$,
\BEqA
t_{\rm min} &=& \frac{c_1}{g}, \nonumber \\
\frac{\Omega_{1,2}}{g} &=& 
\sqrt{
\left(\frac{2\pi n}{c_1}\right)^2-\left(\frac{k-1}{2}\right)^2}\pm 
\sqrt{
\left(\frac{2\pi m}{c_1}\right)^2-\left(\frac{k+1}{2}\right)^2}.
\EEqA
 For [CNOT], $(c_1, c_2, c_3)=(\pi/2, 0, 0)$, which gives
\BEqA
t_{\rm min}&=&\pi/2g, \nonumber \\
\frac{\Omega_{1,2}}{g} &=& 
\sqrt{ 16n^2-\left(\frac{k-1}{2}\right)^2}\pm 
\sqrt{ 16m^2-\left(\frac{k+1}{2}\right)^2}.
\EEqA
Two cases will be distinguished here, depending on the sign of the 
coupling constant.

\subsubsection{$kZZ$ coupling of antiferromagnetic type}

In this case, $k\geq 0$ and $n\geq m = 0,1,2,3,\dots $.
The previous result for $k=0$ is correctly recovered when $n=m=0$.
For $n=m=1,2,\dots$ we get an $(n,n)$-th solution valid in the 
interval $0\leq k \leq 8n-1$. 

Below, we list Rabi frequencies required to generate a {\it perfect 
plug-and-play} [CNOT] in the antiferromagnetic case for several 
values of $n$ and $m$:
\begin{equation}
\label{eq:RABITABLE}
 \begin{array}{|c|cc|cc|cc|cc|}\hline
 k	&
 \Omega^{(1,1)}_1/g & \Omega^{(1,1)}_2/g &
 \Omega^{(2,1)}_1/g & \Omega^{(2,1)}_2/g &
 \Omega^{(2,2)}_1/g & \Omega^{(2,2)}_2/g &
 \Omega^{(3,3)}_1/g & \Omega^{(3,3)}_2/g \\ \hline
0.00	&  7.937254 & 0.000000	&
11.952987 & 4.015733 &
				15.968719 & 0.000000  &
				23.979157 & 0.000000	\\
0.10  & 7.936614  & 0.012600 	&
11.949341 & 4.025327 &
				15.968405 & 0.006262  &
				23.978949 & 0.004170  \\
0.25  & 7.933253  & 0.031513  &
11.942076 & 4.040336 &
				15.966755 & 0.015658	&
				23.977852 & 0.010426  \\
0.50	&	7.921238  & 0.063121	&
11.925151 & 4.067034 &
				15.960859 & 0.031327	&
				23.973935 & 0.020856  \\
1.00  &  7.872983 & 0.127017	&
11.872983 & 4.127017 &
				15.937254 & 0.062746	&
				23.958261 &	0.041739	\\
3.00  &  7.337085 & 0.408882	&
11.401356 & 4.473152 &
				15.683221 & 0.191287	&
				23.790420 &	0.126101	\\ 
5.00  &  6.109853 & 0.818350	&
10.391718 & 5.100215 &
				15.162165 & 0.329768	&
				23.451110 & 0.213209  \\
7.00	&	 2.645751 & 2.645751	&
7.416198 & 7.416198 &
				14.344402 & 0.487995  & 
				22.932659 & 0.305242  \\ \hline
9.00  & 					&						&
& &
				13.173201 & 0.683205	&	
				22.222421 & 0.404996	\\
11.00 & 					&						&
& &
				11.536501 & 0.953495 	&
				21.301017 & 0.516407	\\
13.00 & 					&						&
& &
				9.164486 & 1.418519 	&
				20.139099 & 0.645511 	\\
15.00 & 					&						&
& &
				3.872983	&	3.872983	& 
			 18.691066  &	0.802522	\\ \hline		 
17.00	&	 					&						&
& &
									&						&
				16.881526 & 1.007018	\\		
19.00	& 					&						&
& &
									&						&
				14.570504 & 1.304004  \\
21.00  & 			  	&						&
& &
									&						&
				11.429081 & 1.837418	\\
23.00  &					&						&
& &
									&						&
				4.795832  & 4.795832   \\
\hline
 \end{array}
\end{equation} 
Notice that the product of Rabi frequencies for solutions of 
$(n,n)$-type is always equal to $k$,
\BEq
\Omega^{(n,n)}_1 \Omega^{(n,n)}_2/g^2 = k.
\EEq

\subsubsection{$kZZ$ coupling of ferromagnetic type}

In this case, $k\leq 0$ and $n > m = 0, 1, 2, \dots$.
A particularly interesting family of solutions occurs for $m=0$. 
These solutions ({\it cf.} (\ref{eq:UPTOPHASESOLUTION})) exist 
only for the coupling constant $k=-1$:
\BEq
\Omega^{(n,0)}_{1,2}/g=\sqrt{16n^2-1}, \quad n = 1,2,3,\dots .
\EEq
For certain $n>m = 1, 2, 3$ we find:
\begin{equation}
\label{eq:RABITABLEknegative}
 \begin{array}{|c|cc|cc|cc|}\hline
 k	&
 \Omega^{(2,1)}_1/g & \Omega^{(2,1)}_2/g &
 \Omega^{(3,1)}_1/g & \Omega^{(3,1)}_2/g &
 \Omega^{(3,2)}_1/g & \Omega^{(3,2)}_2/g \\ \hline
-0.00	&  11.952987 & 4.015733&
15.958206 & 8.020952 &
				 19.973939 & 4.005219 	\\
-0.10  & 11.955678 & 4.006464	&
15.961996 & 8.012782 &
				 19.974723 & 4.000055\\
-0.25  &  11.957932 & 3.993165 &
15.966096 & 8.001330 &
				  19.974919 & 3.992507\\
-0.50	&		11.956946 & 3.972586	&
15.968719 & 7.984360 &
				  19.972632 & 3.980447\\
-1.00  &  11.937254 & 3.937254	&
15.958261 & 7.958261 &
					19.958261 & 3.958261\\
-2.00  &  11.826744 & 3.889490	&
15.874508 & 7.937254&
					19.890241 & 3.921521\\ 
-3.00  &  11.618950 & 3.872983	&
15.705143 & 7.959176&
					19.769413 & 3.894906\\ 
-4.00  &  11.307441 & 3.891243 	&
15.444795 & 8.028595 &
					19.594811 & 3.878578\\ 
-5.00  &  10.880300 & 3.952097	&
15.083052 & 8.154848 &
				  19.364917 & 3.872983\\
-6.00	&	 	10.316246 & 4.071248&
14.600739 & 8.355741 &
				  19.077582 & 3.878898\\ 
-7.00	&	 	9.573955  & 4.282452&
13.959460 & 8.667957 &
				  18.729907 & 3.897510\\ 
-8.00	&	 	8.550870 & 4.677887&
13.060789 & 9.187806 &
				  18.318045& 3.930550\\ 
-9.00  & 	6.244998 & 6.244998&
10.908712&10.908712&	
					17.836915& 3.980509\\ \hline
-11.00 & 	&& 	& &
					16.637303 & 4.147307\\
-13.00 & 	&&	& &
					15.038297 & 4.455292\\
-15.00 & 	&&	& &
			   	12.817255 & 5.071289\\	 
-17.00	&	&&	& &
					7.937254 & 7.937254\\		
\hline
 \end{array}
\end{equation} 

\section{Local $rf$-pulses}
\label{sec:rfpulses}

Here we give two (ideal) local $rf$-pulse sequences needed to 
produce the true, canonical CNOT from the [CNOT] located in the 
Weyl chamber via
\BEq
{\rm CNOT} = e^{i\,\pi/4}\, k_1 \, {\rm [CNOT]}\, k_2.
\EEq

\begin{description}

\item[Sequence 1:] Total rotation $\varphi=3\pi/2$,
\BEq
\label{eq:finalrfpulses1}
k_1=e^{-\frac{\pi}{2}\, Y_1}\,
   e^{\frac{\pi}{2}\left(X_1 - X_2\right)},\quad k_2 =
   e^{\frac{\pi}{2}\,Y_1}.
\EEq

\item [Sequence 2:] Total rotation $\varphi=3\pi/2$,
\BEq
\label{eq:finalrfpulses2}
k_1=e^{-\frac{\pi}{2}\,Y_1}, \quad
k_2=e^{-\frac{\pi}{2}\, Z_1}\, e^{\frac{\pi}{2}\left(X_1-X_2\right)}.
\EEq

\end{description}

\section{Fidelity degradation by switching to resonance}
\label{sec:FASTSWITCHING}

We are now going to consider the effect of switching --- the actual 
process of bringing the qubits to resonance with external {\it rf}-field. 

We will assume that [CNOT] generation is performed in three stages:
\begin{description}
\item [Stage 1. Tuning:] Very fast linear rise of Rabi frequencies 
to their optimal values (for a given $k$), with time-dependent Hamiltonian
\BEq
-i H_{\rm tune} = -\left[\left( \Omega^{\rm opt}_1\,X_1
+ \Omega^{\rm opt}_2\,X_2\right)\, \frac{t}{\epsilon\, t_{\rm gate}}
+ g (XX +YY +kZZ) \right],
\EEq
where $\epsilon \ll 1$ characterizes the fraction of the total gate 
time spent in the switching mode (tune/detune). We denote the 
corresponding time evolution by $U_{\rm tune}$;
\item [Stage 2. Resonance:] Evolution with optimal Rabi frequencies 
for most part of [CNOT]
generation, $U_{\rm resonance}$, where
\BEqA
U_{\rm resonance} &=& e^{-i\,(1-2\,\epsilon)\,t_{\rm gate}\, 
H_{\rm resonance}},
\\ \nonumber
-i H_{\rm resonance} &=& -\left[\Omega^{\rm opt}_1\,X_1
+ \Omega^{\rm opt}_2\,X_2 + g (XX +YY +kZZ) \right],
\EEqA
and
\item [Stage 3. Detuning:] Fast linear decrease of Rabi frequencies 
to zero, with time-dependent Hamiltonian
\BEq
-i H_{\rm detune} = -\left[\left( \Omega^{\rm opt}_1\,X_1
+ \Omega^{\rm opt}_2\,X_2\right)\,\left( 1- \frac{t}{\epsilon\, 
t_{\rm gate}}\right)
+ g (XX +YY +kZZ) \right].
\EEq
\end{description}

The full gate is then the product
\BEq
\label{eq:OPTIMALCNOT}
{\rm [CNOT]}_{\rm opt} =  U_{\rm detune} U_{\rm resonance} U_{\rm tune}.
\EEq

In order to find the optimal values of the gate parameters, we 
first truncate the Magnus expansion for each switching part of 
the gate:
\BEq
U_{\rm switching}\left(\epsilon \, t_{\rm gate}\right) =
e^{\sigma \left(\epsilon \,t_{\rm gate}\right)},
\EEq
with
\BEqA
\s \left(\epsilon \,t_{\rm gate}\right)&\approx& 
-i \int_0^{\epsilon \,t_{\rm gate}} dt  H_{\rm switching}(t)\\ \nonumber
&& \quad + \frac{(-i)^2}{2}\int_0^{\epsilon \,t_{\rm gate}}dt_1
\left[ H_{\rm switching}(t_1), \, 
\int_0^{t_1} dt_2\,
 H_{\rm switching}(t_2) \right],
\EEqA
and then use the 4th order iteration method \cite{ISERLES99}
\BEqA
U_{n+1}& = &
e^{\sigma_n}\,U_n, \\ \nonumber
\s_n &=& \frac{h}{2}\, (A_1 + A_2) + \frac{\sqrt{3}\, h^2}{12}\,
[A_1,\, A_2], \\ \nonumber
A_1 &=& -i H_{\rm switching}(t_n + c_1h), \\ \nonumber
A_2 &=& -i H_{\rm switching}(t_n + c_2h), \\ \nonumber
h &=& \frac{\epsilon \,t_{\rm gate}}{N},
\EEqA
where $N$ is the number of iteration steps and
\BEq
c_{1,2} = \frac{1}{2} \pm \frac{\sqrt{6}}{3} 
\EEq
are the nodes of the Gauss-Legandre 4th order quadrature in 
$[0,\, 1]$. As described in \cite{ISERLES99}, despite the 
presence of only $h$ and $h^2$ terms, this is indeed the 4th 
order approximation to the Magnus expansion.

Optimization of gate's parameters was performed using Nelder-Mead 
simplex direct search with bound constraints \cite{DERRICO} for the 
minimum of the square of the Frobenius distance
\BEq
d^2_{\rm Frobenius} = {\rm tr}\left[\left({\rm [CNOT]}-
{\rm [CNOT]}_{\rm opt}\right)^\adj
\left({\rm [CNOT]}-{\rm [CNOT]}_{\rm opt}\right)\right]
\EEq
between the optimal ${\rm [CNOT]}_{\rm opt}$ given by 
(\ref{eq:OPTIMALCNOT}) and the perfect [CNOT] given in 
(\ref{eq:CNOTMATRIX}). Due to the enormous richness of 
possible CNOTs, we optimized only the most interesting of them. 
With $N = 2500$ we simulated the switching mechanism for 
$\epsilon = 0.025$:
\begin{equation}
 \begin{array}{|c|c|c|c|c|}\hline
k  	&	 
\Omega^{\rm opt}_1/g & 
\Omega^{\rm opt}_2/g & 
t_{\rm gate}, \times \pi/(2g) &
d^2_{\rm Frobenius}\,, \, \times 10^{-3}\\ \hline 
0.50  &  	 8.130446 &  0.064667 &  0.999268 & 1.9	  \\ 
0.25  &  8.141971&   0.032287 &  0.999363 & 1.6	 \\ 
0.10  & 8.145193&  0.012910 &  0.999390	&	1.5	\\ 
0.00	&	8.145807  	&  	0.000000 		&	0.999395 &	1.5  \\ \hline 
-0.10 &  12.269887&   4.111716&  0.999382&1.5	 \\ 
-0.25	&	  12.272602 &   4.098223 & 0.999348	& 1.6	   \\
-0.50 & 	12.272970&   4.077597&   0.999237	&	1.9  \\ 
 \hline
 \end{array}\quad 
\end{equation} 

Using the local pulses given in Section \ref{sec:rfpulses} 
we can transform the corresponding optimal 
${\rm [CNOT]}_{\rm opt}$ to its canonical form. For example,
\BEq
\label{eq:CNOTMATRIXOPTIMAL0}
{\rm CNOT}_{k=0.25} =
\left[
\begin{array}{cccc}
 0.9998 - 0.0025i & -0.0001 - 0.0017i & -0.0001 - 0.0047i  & 0.0192i\\
  -0.0001 - 0.0017i  & 0.9998 - 0.0025i  &- 0.0193i  & 0.0001 + 0.0047i\\
   0.0193i   &0.0047i &  0.0001 - 0.0007i  & 0.9998 + 0.0025i\\
  -0.0001 - 0.0047i   & - 0.0192i &  0.9998 + 0.0025i  & 0.0001 - 0.0007i\\  
  \end{array}
\right],
\EEq
whose intrinsic fidelity relative to the computational basis is
\BEq
F = \frac{1}{4}\sum_{r=1}^4 \left| \left({\rm CNOT}^\adj 
\cdot {\rm CNOT}_{k=0.25}\right)_{rr}\right| =0.9998.
\EEq

\section{Conclusion}

To recapitulate, we have demonstrated how to implement various 
high-fidelity, plug-and-play CNOT logic gates by a single 
application of Hamiltonians used to describe coupled superconducting 
qubits. Architectures involving capacitive and inductive couplings 
have been analyzed for which the physical parameters required to 
generate the perfect CNOT have been found in closed, analytic form. 
Additional numerical simulations based on 4th order Magnus expansion 
were used to correct this ideal limit by taking into account 
the switching to resonance.


\section*{Acknowledgments}

This work was supported by the Disruptive Technology Office under
grant W911NF-04-1-0204 and by the National Science Foundation under
grant CMS-0404031.

The author would like to thank Michael Geller, Emily Pritchett, 
and Andrew Sornborger for helpful discussions.

 \renewcommand{\theequation}{A-\arabic{equation}}
  \setcounter{equation}{0}  
  \section*{Appendix: Derivation of the Hamiltonian for inductively 
  coupled flux qubits} 

The system consists of two superconducting loops of self-inductance 
$L$ coupled by a mutual inductance $M$ and driven by external 
magnetic fluxes $\Phi_i$, as shown on Fig. 1. 
Each loop is interrupted by a Josephson junction of critical current  
$I_0$ and effective capacitance $C$.

 Inside the superconducting material, the Ginzburg-Landau complex order 
 parameter can be written in the usual form,
\BEq
\psi({\bf r})=\sqrt{\rho_{\rm s}({\bf r})}e^{i\theta({\bf r})},
\EEq
where $\rho_{\rm s}$ is the density of Cooper pairs. The supercurrent 
flux is
\BEq
{\bf j}_{\rm s} = \frac{\hbar \rho_{\rm s}}{m^{\ast}}\left(\nabla \theta 
- \frac{q}{\hbar}{\bf A}_{\rm s}\right),
\EEq
where ${\bf A}_{\rm s}$ is the vector potential inside the superconductor 
and $q=-2e$. The gauge is assumed to be chosen in such a way as to 
guarantee $\bf \hat  n \cdot  j_{\rm s} = 0$ at the surface. In the clean 
limit, $\bf j_{\rm s}=0$ inside the material, which leads to
\BEq
{\bf A}_{\rm s}= \frac{\hbar}{q}\nabla \theta .
\EEq

The line integral of $\bf A$ around Loop 1 is (here, the limits of integration 
1 and 2 denote the two sides of the junction, the rest of notation is 
clear from context):
\BEqA
\label{eq:LOOPINTEGRAL1}
\oint_{\rm Loop\; 1} {\bf A}d{\bf l} &=& 
\left(\int^2_1 {\bf A}d{\bf l}\right)_{\rm inside\; JJ}
+\left(\int_2^1 {\bf A}d{\bf l}\right)_{\rm in\; the\; rest\;of\;circuit\;
( inside\;superconducting\; bulk)}
\nonumber \\
&=&
\left[\int_1^2 {\bf A}_{\rm JJ}d{\bf l} -\frac{\hbar}{q}(\theta_2-\theta_1)\right]
+ \left[ \frac{\hbar}{q}(\theta_2-\theta_1) -\int^2_1 {\bf A}_{\rm s}d{\bf l}\right]
\nonumber \\
&=&
-\frac{\hbar}{q} \left[(\theta_2-\theta_1)-
\frac{q}{\hbar}\int_1^2 {\bf A}_{\rm JJ}d{\bf l}\right]
- \frac{\hbar}{q}\left[ (\theta_1-\theta_2) + \int^2_1 \nabla \theta
d{\bf l}\right]
\nonumber \\ 
&=&
-\frac{\hbar}{q} 
\left[
\phi_1 + \oint_{\rm Loop\; 1} \nabla \theta
d{\bf l}\right]
\nonumber \\
&=& -\frac{\hbar}{q} (\phi_1 + 2\pi n_1),
\EEqA
where $\phi_1$ is the gauge invariant phase difference across the first 
junction, and $n_1=0,1,2,3,\dots$. This ``quantization'' condition is 
justified by the fact that the order parameter is nonzero inside the 
junction \cite{JACOBSON65}. Alternatively,
\BEq
\label{eq:LOOPINTEGRAL2}
\oint_{\rm Loop\; 1} {\bf A}d{\bf l}
=
\Phi^{\rm external}_1+\Phi^{\rm self}_{1}+\Phi^{\rm mutual}_{1}
= 
\Phi_1-LI_1 + MI_2.
\EEq

Introducing
\BEq
\a = \frac{\hbar}{2e}=\frac{\Phi_{\rm sc}}{2\pi}, \quad 
E_J = \a I_{0}, \quad \omega_0 = \frac{1}{\sqrt{LC}}, \quad 
E_0= \frac{\a^2}{L}=
\frac{\hbar^2\omega_0^2}{2E_C}, 
\quad E_C = \frac{(2e)^2}{2C},
\quad \Upsilon = \frac{M}{L},
\EEq
and taking into account (\ref{eq:LOOPINTEGRAL1}) and 
(\ref{eq:LOOPINTEGRAL2}) together with the Josephson equations,
we get the equations of motion for the two gauge-invariant phase 
differences,
\BEqA
\label{eq:EQSOFMOTIONBAD}
\a^2 C\left(\ddot{\phi_1}-\Upsilon \ddot{\phi_2}\right) 
&=& -E_J\left(\sin \phi_1-\Upsilon \sin \phi_2\right)
-E_0
\left(\phi_1 + 2\pi n_1 
-\frac{2\pi\Phi_1}{\Phi_{\rm sc}}\right),
\nonumber \\
\a^2 C\left(\ddot{\phi_2}-\Upsilon\ddot{\phi_1}\right) 
&=& -E_J\left(\sin \phi_2-\Upsilon\sin \phi_1\right)
-E_0
\left(\phi_2 + 2\pi n_2 
-\frac{2\pi\Phi_2}{\Phi_{\rm sc}}\right).
\EEqA
It is clear that $2\pi n_i$ are not dynamical and can be absorbed into 
$\phi_i +2\pi n_i \rightarrow \phi_i$
without affecting any physics. Notice also that in the limit 
$M\rightarrow 0$ we correctly recover a well-known result for two 
independent loops.

Because of the presence of terms $\Upsilon E_J\sin \phi_i$ in 
(\ref{eq:EQSOFMOTIONBAD}), we cannot immediately deduce system's 
Lagrangian. However, by multiplying the second equation in 
(\ref{eq:EQSOFMOTIONBAD}) by $\Upsilon$, and by adding the result to 
the first 
equation, we can separate the variables
\BEq
\a^2 C\left(1-\Upsilon^2\right) \ddot{\phi_1}
= -E_J\left(1-\Upsilon^2\right) \sin \phi_1
-E_0
\left[\left(\phi_1 
-\frac{2\pi\Phi_1}{\Phi_{\rm sc}}\right)
+\Upsilon\left(\phi_2 
-\frac{2\pi\Phi_2}{\Phi_{\rm sc}}\right)
\right],
\EEq
and similarly for $\ddot{\phi_2}$. The Lagrangian is now easily 
found to be
\BEqA
L &=& \left(1-\Upsilon^2\right) 
\left\{
\frac{\a^2 C}{2} \left(\dot{\phi_1}^2+\dot{\phi_2}^2\right) 
+ E_J (\cos \phi_1+\cos \phi_2)
\right\}
\nonumber \\
&& \quad \quad \quad \quad \quad \quad
- \frac{E_0}{2}
\left[\left(\phi_1 
-\frac{2\pi\Phi_1}{\Phi_{\rm sc}}\right)^2
+\left(\phi_2 
-\frac{2\pi\Phi_2}{\Phi_{\rm sc}}\right)^2\right]
\nonumber \\
&& \quad \quad \quad \quad \quad \quad
 \quad \quad \quad \quad \quad \quad
- \Upsilon E_0
\left(\phi_1  
-\frac{2\pi\Phi_1}{\Phi_{\rm sc}}\right)
\left(\phi_2  
-\frac{2\pi\Phi_2}{\Phi_{\rm sc}}\right).
\EEqA
In the limit of weak coupling ($\Upsilon \ll 1$) the terms 
$\sim \Upsilon^2$ 
can be dropped, giving
\BEq
H = \frac{p_1^2+p_2^2}{2m}+U(\phi_1, \phi_2),
\EEq
with $p_i = m\dot{\phi}_i$, $m = \a^2C$, and
\BEq
U(\phi_1, \phi_2) = 
\sum_{i=1}^{2} \left[
-E_J \cos \phi_i + \frac{E_0}{2}
\left(\phi_i -\frac{2\pi\Phi_i}{\Phi_{\rm sc}}\right)^2\right]
+ \Upsilon E_0
\left(\phi_1 
-\frac{2\pi\Phi_1}{\Phi_{\rm sc}}\right)
\left(\phi_2  
-\frac{2\pi\Phi_2}{\Phi_{\rm sc}}\right).
\EEq
After projecting onto the qubit subspace and using RWA we recover 
(\ref{eq:tildeH}).

\end{document}